# An Indoor Tracking System Based on Bluetooth Technology

Samuel King Opoku, *Member, IEEE*

*Abstract*—Implementations of tracking systems have become prevalent issues in modern technology due to its advantage of location detection of objects. Objects are usually tracked using trackers based on GPS, GSM, RFID and Bluetooth signal strength implementation. These mechanisms usually require line-of-sight operations, limited coverage and low-level programming language for accessing Bluetooth signal strength. This paper presents an alternative technique for tracking the movement of indoor objects based on Bluetooth communication technology, principles of motion and least square statistical method. Algorithms are designed and implemented using Java.

## I. Introduction

TRACKING systems, systems designed to monitor devices or persons, have become prevalent issues in modern technology. There are many advantages of locating and tracking a person or object, in a variety of contexts such as following the movements of a child around an amusement park, locating colleagues in an office or tracking the movement of luggage through an airport. The predominant mechanisms for tracking humans involve the use of video surveillance systems. These systems require human operator to monitor the CCTV images at a central location. Loss of concentration usually occurs when fatigue sets in [1]. Vehicles and other objects are usually tracked using trackers whose implementation is based on Global Positioning System (GPS). These systems display the location of a vehicle within a specified time frame. GPS, however, supports outdoor navigation since it requires lines-of-sight operation with at least three satellites [2]. Another technique for implementing tracking system is by using Radio Frequency IDentification (RFID). RFID uses either passive or active tags to track objects [3]. Passive RFID tracking is very common in shops and libraries where tags are attached to products and are checked as they leave the shop by passing through receivers near the doors. Active RFID is popularly used in warehouses and locations like airports where a larger range is needed. RFID tracking uses ultra-low power and there is no need for line-of-sight operation. While RFID tags are very cheap, small and suitable for tracking objects, the sensors are considerably more expensive and require extensive configuration and software installations [3], [4]. RFID signals are easily blocked by objects and other radio waves [4]. One more method for tracking objects is based on GSM communication technology. The GSM equipment communicates with the GSM network through relay stations. The times at which signal arrive together with the angle of arrival from at least three stations allow location detection through triangulation [5]. The main problem with GSM is inaccuracy in location determination due to its limited coverage in densely populated area [4]-[6].

With the range of personal devices using Bluetooth, the possibility arises to locate and track the movements of objects. Bluetooth has become an emerging technology for determining indoor and sometimes outdoor position of a communicating device. Although there is no specific support for positioning service in Bluetooth technology yet the predominant technology used are signal strength measurement, link quality and bit error rate which rely on the services of the Host Controller Interface [7], [8]. Thus the Received Signal Strength Indicator (RSSI) value of the Bluetooth protocol is used to get a correlation to the distance between sender and receiver in a network. The RSSI value in providing the distance between the received signal strength and an optimal received power rank is called the Golden Receiver Power Rank (GRPR) [9]. If the value of RSSI is in GRPR defined by zero, no unique function can be approximated [8], [9]. This paper presents an alternative technique to track the movement of indoor objects based on Bluetooth technology, least square statistical method and principles of motion. The device being tracked is connected to three Access Points and a Central Monitoring System. PC's with Bluetooth radio adaptors which run multi-threaded Java based desktop applications are used to implement the Access Points and the Central Monitoring System. The device being tracked runs a J2ME application.

The Bluetooth communication protocol has client-server architecture. The client initiates the connection and the server accepts or receives the connection. Bluetooth Specification [10] consists of Bluetooth protocol stack and profiles. The protocol stack is a software that has direct access to the Bluetooth device controlling device settings, communication parameters and power levels for the Bluetooth device. The main implemented layers in the stack are the Host Controller Interface (HCI), Logical Link Controller Adaptation Protocol (L2CAP), Service Discovery Protocol (SDP) and Object Exchange Protocol (OBEX) [8], [10]. Communication between devices depends on the type of data transferred.

OBEX protocol supports exchange of such physical data as files and images. L2CAP is used for sending packets between host and client whereas Radio Frequency COMMunication (RFCOMM) supports simple data transfer [10]-[12]. The Bluetooth profiles allow different Bluetooth devices to interoperate. Although RFCOMM is easy to setup yet a one-way communication link usually shuts down before data is transferred [10]. This problem is addressed by allowing bidirectional transmission before shutting down the link. The range of coverage of Bluetooth depends on the class of the Bluetooth adaptor [12]. This could be class 1 type which covers up to 100 meters, class 2 type covers up to 10 meters and class 3 type usually covers up to 1 meter. Virtually all devices use class 2, with class 3 being adopted for some smaller devices such as Bluetooth telephony headsets. Bluetooth is also limited by excessive power consumption [8] and the power level of the energy source [13]. The figure below provides the general overview of the Bluetooth architecture.

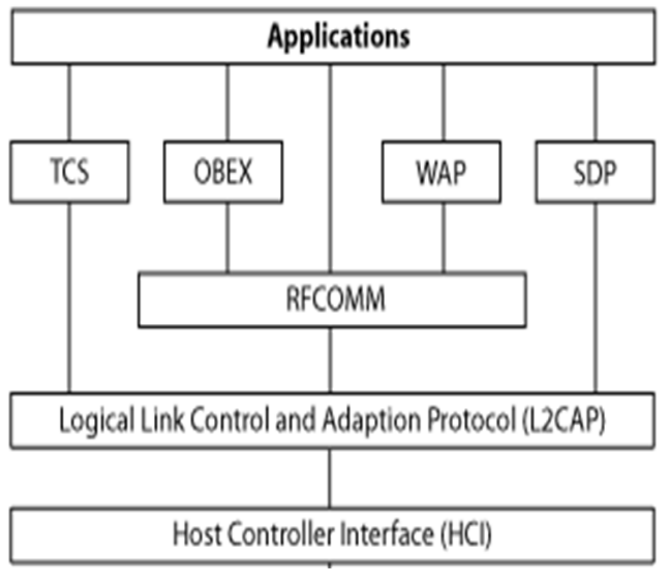

Fig. 1. Overview of Bluetooth Architecture

J2ME is the platform used in order to develop applications for portable devices. J2ME is made up of two configurations. These are Connected Limited Device Configuration (CLDC) and Connected Device Configuration (CDC) [14]. The application for mobile phones is developed using MIDlet with the help of CLDC configuration supported by MIDP profile [15]. The CDC configuration has three profiles. The first type is the Foundation Profile (FP) which is a set of Java API for devices having limited resources that do not need a graphical user interface system [16]. It is usually used in embedded devices. It is specified in JSR 219 [17]. The second type is the Personal Basic Profile (PBP) which is a superset of the foundation profile. This profile supports devices with lightweight graphical user interface requirement [18]. Its specification is described in JSR 217 [19]. The last type is the Personal Profile (PP) which is an extension of the Personal Basic Profile with a graphical user interface toolkit based on AWT [20]. It is intended for such higher devices as PDA's, smart communications, game consoles and automobile dashboard. Its specification is described in JSR 62 [21]. The Bluetooth API (JSR 82) is integrated with the above-mentioned APIs when Bluetooth is required. The compatibility of J2ME and J2SE enables the development of a backend server to assist the limited processing power of mobile devices. J2SE is ideally used for GUI applications [22]. The Bluecove API is integrated with J2SE for Bluetooth communication between the backend server and the mobile devices [11], [23].

## II. SYSTEM ARCHITECTURE

The system is made up of a central monitoring system, the device being tracked and Bluetooth Access Points. The Access Points, in this work, are implemented applications on PCs that have Bluetooth radio adaptors connected to them. Access Points only resend the message (or signal) sent by the device being tracked. The Access Points are placed in the room such that their positions are known in the XY plane. The Central Monitoring system knows the initial location of all the devices being tracked. When the device moves considerably, the monitoring system sounds alarm displaying the new location, in coordinate form, of the device. Each device being tracked is connected to three Access Points. A Bluetooth-enabled tracker is attached to the device being tracked. The tracker communicates with the Central Monitoring System (CMS) and three selected Access Points.

The general overview of the system is illustrated in the figure below:

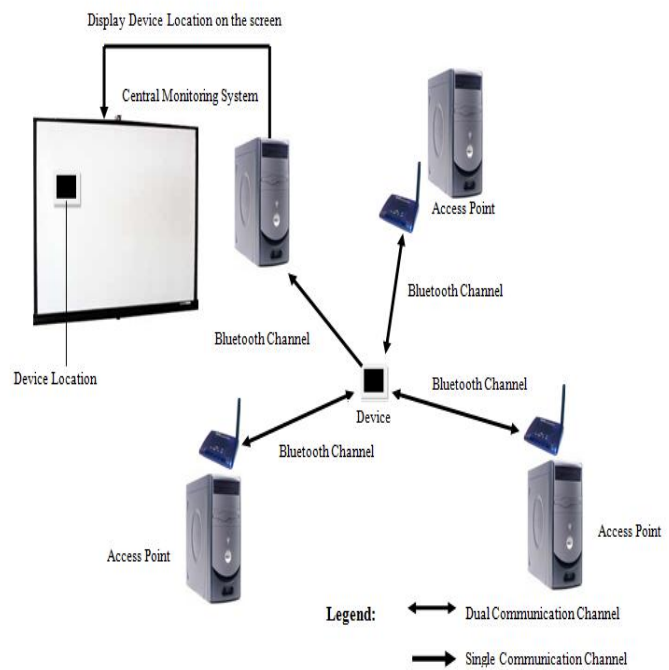

Fig. 2. Overview of System Architecture

## III. TRACKING SYSTEM ALGORITHMS

For simplicity sake, only one device is considered. The device being tracked is connected to three Access Points(AP) whose positions are known in the XY plane as shown below:

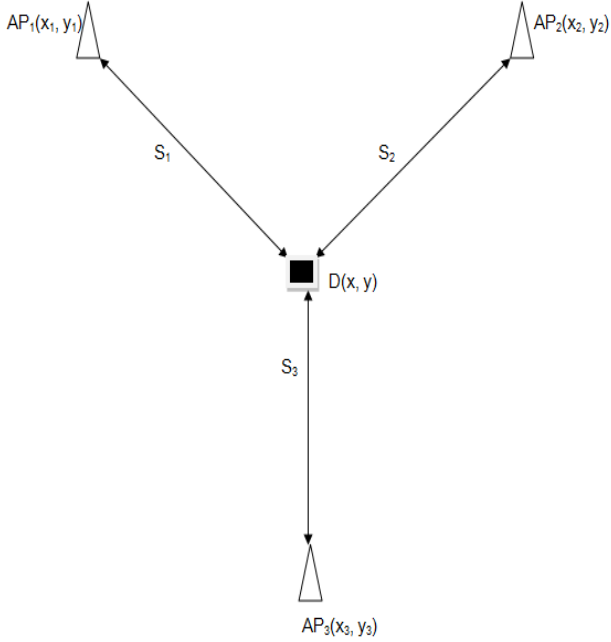

Fig. 3. System Architecture in XY Plane

From Fig. 2, $AP_i$'s are the Bluetooth Access Points, D denotes the device being tracked and $S_i$'s represent the distance between $AP_i$'s and D. All computations are done by the Central Monitoring System.

### A. Device and Access Point Distance Algorithm

The distance between an Access Point (AP) and the device being tracked is based on the principle of motion and least square statistical method. By definition of speed which is the time rate of change of distance which is mathematically expressed as $\text{Speed}(v) = \frac{\text{distance (s)}}{\text{time (t)}}$. It can be written as $V = \frac{dS}{dt}$ and then $dS = Vdt$. Applying the concept of integration, $\int dS = \int V dt$, the resulting equation becomes $S = Vt + C$ with C catering for transmission errors. Given indoor physical conditions, a linear regression of S on T based on least square method determines parameters V and C by respectively finding the gradient and S-intercept. T is the average time taken by a unique signal send from the device being tracked to the Access Point to return back to the device being tracked (that is $T = \frac{\text{total time (t)}}{2}$). Thus the above equation finally becomes $S_i = VT_i + C \ \forall i = 1,2,\ldots,N$ where N is the total number of samples taken in the experiment such that $N \geq 5$

The algorithm used to determine the Signal Speed (V) and the transmission error (C) is described below:
1. Computer average time (T) for each pair as $(T_i, S_i) \ \forall i = 1,2,\ldots,N$
2. Find the mean time and mean distance ($M_T$ and $M_S$) as $M_T = (\sum_{i=1}^{N} T_i)/N$ and $M_S = (\sum_{i=1}^{N} S_i)/N$
3. Determine the deviation for each time and distance as $D_{Ti} = T_i - M_T$ and $D_{Si} = S_i - M_S \ \forall i = 1,2,\ldots,N$
4. Calculate the mean of squares of the time deviations as $MS_T = (\sum_{i=1}^{N} D_{Ti}^2)/N$
5. Calculate the mean of the products of time and distance deviations as $MS_{ST} = (\sum_{i=1}^{N}(D_{Ti} \times D_{Si}))/N$
6. Find the Signal Speed (in meter per second) which represents the regression coefficient as $V = \frac{MS_{ST}}{MS_T}$
7. Compute the transmission error as $C = M_S - M_T \times \frac{MS_{ST}}{MS_T}$

Using the distance between the Access Points and the device being tracked alone usually results inaccurate estimations. Signal attenuation affects transmission time. Thus, to minimize the degree of error, the coordinate of the device being tracked is calculated. When there is a significant change is any of the x or y coordinate then the device has moved.

### B. Algorithm for the Coordinates of the Device Being Tracked

Given three AP's in a communication network and the positions of the AP's, defined by $\vec{P_k} = (x_k, y_k)^k \ \forall k \in 1,2,3$, it follows that the known distances $S_i \ \forall i \in 1,2,3$ between AP and a point in the XY plane which corresponds to the position of the device being tracked represented as $\vec{D} = (x, y)$ is given, based on the Euclidean inner product [24], by $S_i = \sqrt{(x - x_i)^2 + (y - y_i)^2} \ \forall i \in 1,2,3$. This is also written as $S_i^2 = (x - x_i)^2 + (y - y_i)^2 \ \forall i \in 1,2,3$. Three equations are formed from $S_i \ \forall i \in 1,2,3$ such that

$S_1^2 = x^2 - 2x_1 x + x_1^2 + y^2 - 2y_1 y + y_1^2$ -------(1)
$S_2^2 = x^2 - 2x_2 x + x_2^2 + y^2 - 2y_2 y + y_2^2$ -------(2)
$S_3^2 = x^2 - 2x_3 x + x_3^2 + y^2 - 2y_3 y + y_3^2$ -------(3)

The three equations above represent three circles whose radii are $S_i \ \forall i \in 1,2,3$ as shown in the figure below:

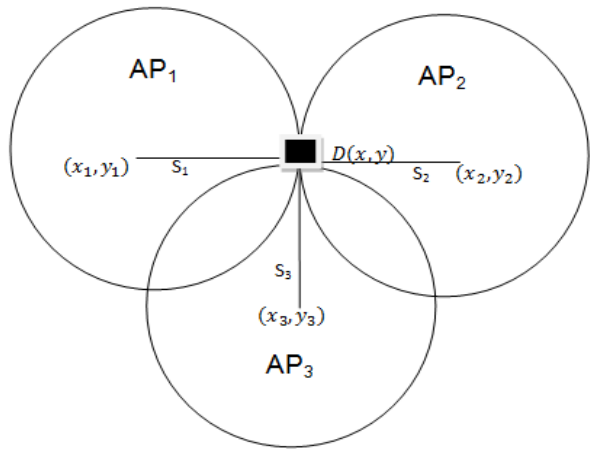

Fig. 4. Geometric Representation of System Architecture

Solving (1) – (2) and (2) – (3), the above equations reduced to two equations as:
$2(x_2 - x_1)x + 2(y_2 - y_1)y = S_1^2 - S_2^2 - x_1^2 + x_2^2 - y_1^2 + y_2^2$
and
$2(x_3 - x_2)x + 2(y_3 - y_2)y = S_2^2 - S_3^2 - x_2^2 + x_3^2 - y_2^2 + y_3^2$

Finally both equations are written in a compact form (representing the equation of a line) as

$$ax + by = e \quad \text{-------------(4)}$$
$$cx + dy = f \quad \text{---------------(5)}$$

where a, b, c, d, e and f are constants and their values are determined as follow: $a = 2(x_2 - x_1)$, $b = 2(y_2 - y_1)$, $c = 2(x_3 - x_2)$, $d = 2(y_3 - y_2)$, $e = S_1^2 - S_2^2 - x_1^2 + x_2^2 - y_1^2 + y_2^2$ and $f = S_2^2 - S_3^2 - x_2^2 + x_3^2 - y_2^2 + y_3^2$

"Given a set of data, a line that will be as closely as possible which approximate each of the data points can be found such that $A\vec{x} = b$ [24]". A is n × m matrix with n equations and m unknowns. Representing equations (4) and (5) in the form $A\vec{x} = b$, it follows that $A = \begin{bmatrix} a & b \\ c & d \end{bmatrix}$, $\vec{x} = \begin{pmatrix} x \\ y \end{pmatrix}$ and $b = \begin{pmatrix} e \\ f \end{pmatrix}$ resulting $\begin{bmatrix} a & b \\ c & d \end{bmatrix} \begin{pmatrix} x \\ y \end{pmatrix} = \begin{pmatrix} e \\ f \end{pmatrix}$ -------------(6)

"Suppose that A is n × m matrix with linearly independent columns, then $A^T A$ is an invertible matrix [24]." Since $A\vec{x} = b$ is equivalent to $A^T A\vec{x} = A^T b$ and $(A^T A)^{-1}$ exists as indicated by [24], it implies that there is a solution for the following equation

$$\vec{x} = (A^T A)^{-1} A^T b \quad \text{------------ (7)}$$

Substituting the values, equation (7) becomes:

$D(x, y) = \vec{x}^T = \left( \begin{bmatrix} a & b \\ c & d \end{bmatrix}^T \times \begin{bmatrix} a & b \\ c & d \end{bmatrix} \right)^{-1} \times \begin{bmatrix} a & b \\ c & d \end{bmatrix}^T \times \begin{pmatrix} e \\ f \end{pmatrix}$. This is simplified as

$$\left( \frac{1}{(a^2 + c^2)(b^2 + d^2) - (ab + cd)^2} \right) \begin{bmatrix} (b^2 + d^2)(ae + cf) - (ab + cd)(be + df) \\ (a^2 + c^2)(be + df) - (ab + cd)(ae + cf) \end{bmatrix}$$

Finally the coordinate of the device being tracked is:

$$x = \frac{(b^2 + d^2)(ae + cf) - (ab + cd)(be + df)}{(a^2 + c^2)(b^2 + d^2) - (ab + cd)^2}$$

$$y = \frac{(a^2 + c^2)(be + df) - (ab + cd)(ae + cf)}{(a^2 + c^2)(b^2 + d^2) - (ab + cd)^2}$$

C. *Algorithm for Triggering Device Movement Alarm*

A fractional change in either x-coordinate or y-coordinate is ignored. This ensures that there is a significant change in position before the alarm sounds. The flowchart in Fig. 5 shows the algorithm for monitoring the device being tracked to determine significant movement.

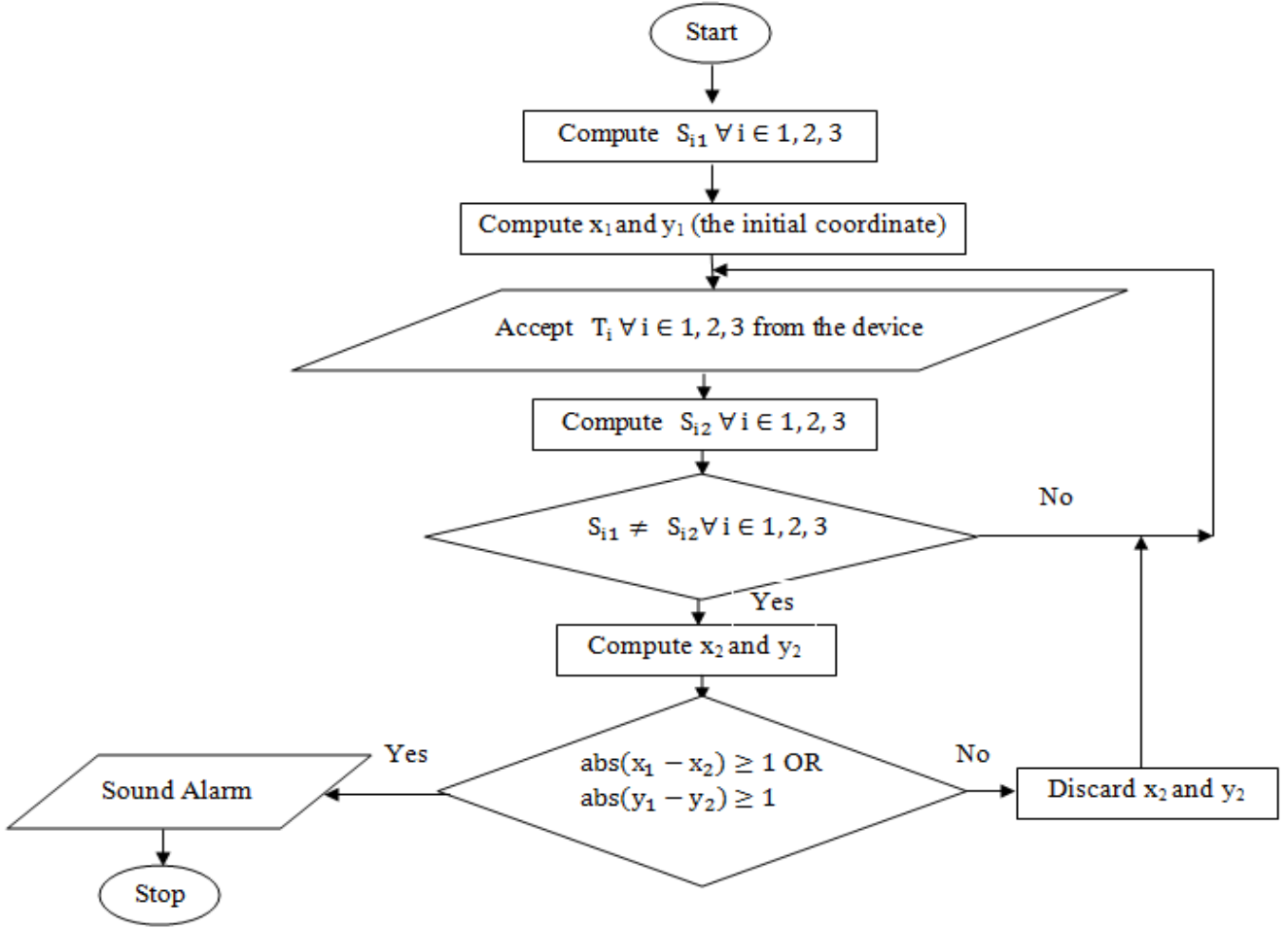

Fig. 5. Alarm Triggering Process

## IV. SYSTEM IMPLEMENTATION

The Access Points and the Central Monitoring System applications were implemented using J2SE. The Bluecove API was used to communicate with the device being tracked. The application for the tracker which is attached to the device being tracked was implemented in J2ME. Devices implemented as trackers do not have graphical user interface. However, devices used in the quantitative analysis to determine signal speed (V) and transmission error (C) have simple graphical user interface. Access Points resend the signal sent by the device being tracked. The time associated with the transmission is sent by the device being tracked to the Central Monitoring system which uses the time to determine the distance between the Access Points and the device being tracked. With these distances, the coordinate of the device being tracked is calculated. The Access Points and the Central Monitoring System have different Universally Unique Identifiers (UUID) indicating different services provided by each. In this implementation, the trackers were attached to objects like luggage and chair

### A. Implementation of Access Point System

There is no graphical user interface associated with this implementation. For large-rang radius coverage, the Bluetooth radio adaptor used was class 1 type with a coverage of up to 100m. Each Access Point system has unique UUID and a three-letter code which is assigned at the system's console when it is started. The figure below illustrates the configuration of an Access Point.

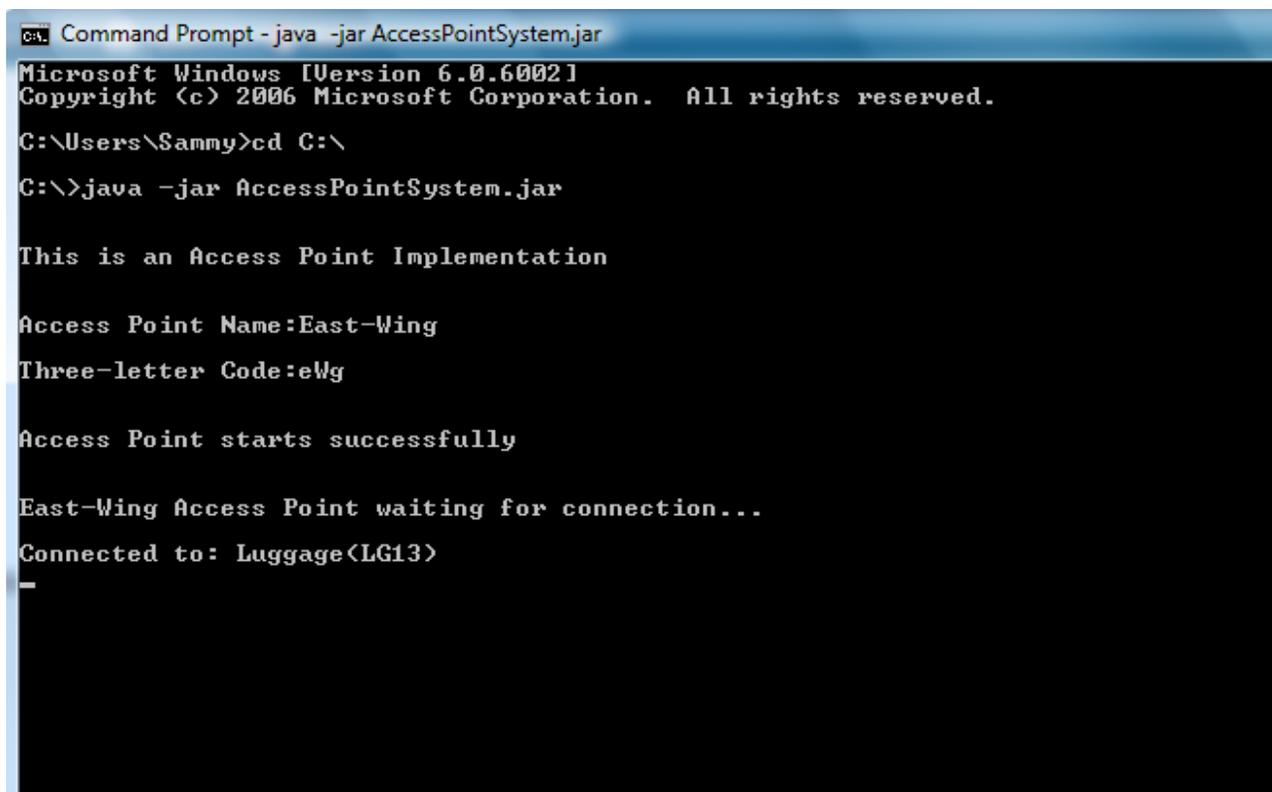

Fig. 6. Access Point Configuration

Once there is a connection between the device being tracked and the Access Point, the Access Point sends its code which is appended to the signal to be used for device tracking. The signal is the same for all devices except that is has the three-letter code of the Access Point appended to it. The Access Point is a multithreaded application and it can connect to other devices being tracked at the same time. However, it uses the three-letter code to communicate with all devices connected to it.

### B. Implementation of Device Being Tracked

There are two categories of device implementations. The first category is used to determine transmission parameters (signal speed and transmission error). This category has a graphical user interface for sending signals to a connected Access Point and display the time of transmission.

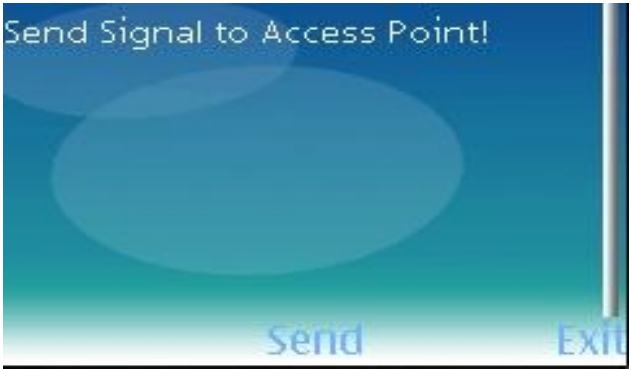

Fig. 7. Parameters Determination Interface

The other category of the implementation is for the tracker systems. This type of implementation has no graphical user interface. When the system is started, the tracker attached to the device being tracked automatically connects to three Access Points and the Central Monitoring System. When there is an error, a message is displayed and the tracker shuts down. The device is restarted to make a reconnection attempt. The system is implemented such that three Access Points and the Central Monitoring System should be selected and connected before tracking can be done.

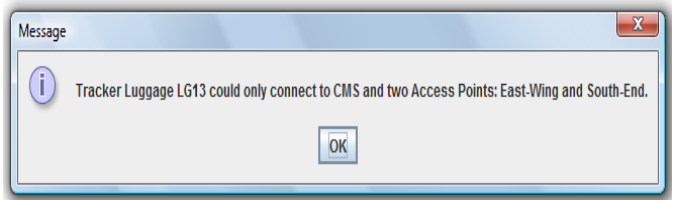

Fig. 8. Error Message during Device Connectivity

### C. Signal Speed and Transmission Error Determination

Manually measure at least five distances (in meters) between an Access Point and a device being tracked. A message is sent, for each distance measured, from the device to the Access Point and the time taken by the message to return is recorded. An interface is provided by the Central Monitoring System to capture each distance and the corresponding time. The interface is illustrated in the figure below:

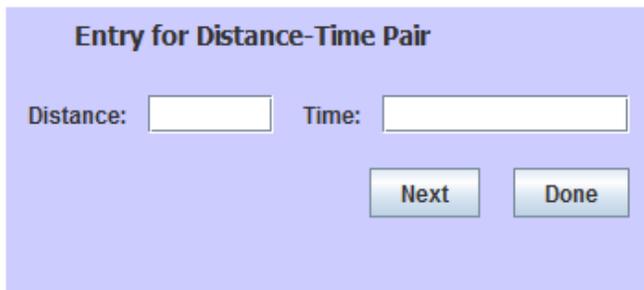

Fig. 9. Interface for Distance-Time Pair Entry

### D. Implementation of Central Monitoring System

The Central Monitoring System, a multithreaded application, only connects to the devices being tracked. There is no connection between the Central Monitoring System and the Access Points. The Central Monitoring System only needs the coordinates of the Access Points. The Central Monitoring System basically performs two functions. One of the functions is the initialization of the system which involves determination of signal speed and transmission error by clicking on "Initialization" button. Any time the Initialization button is clicked, Fig. 9 is displayed so that the user enters the distance-time pairs. When "Done" button is clicked, the signal speed and the transmission error are calculated. When the "Done" button is clicked without five or more entries of distance-time pairs, the user is prompted. The user has the chance to continue with the entry or abort the whole initialization process. Users are also allowed to enter the coordinates of the Access Points by clicking on "Access Point Coordinates" button. This is illustrated in the figure below:

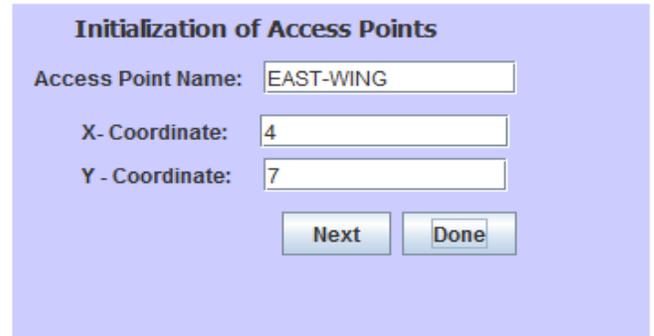

Fig. 10. Interface for Access Points Initialization

The other function of the Central Monitoring System is to monitor the movement of objects. This is the default function of the system. In tracking an object, each signal sent to the Central Monitoring System has unique identification which determines the source of the message. The signal is formatted as: SourceID,time_AP1,time_AP2,time_AP3. This signal is transmitted by devices being tracked every five seconds. The first transmission from every device after system initialization (that is determination of parameters – signal speed and transmission error) is interpreted as the initial location of the device. The name and the ID of the device being tracked form the user friendly name of the Bluetooth System. The user friendly name can be modified since a tracker used for one device can be used for on another device. The figure below displays the initial location of some devices

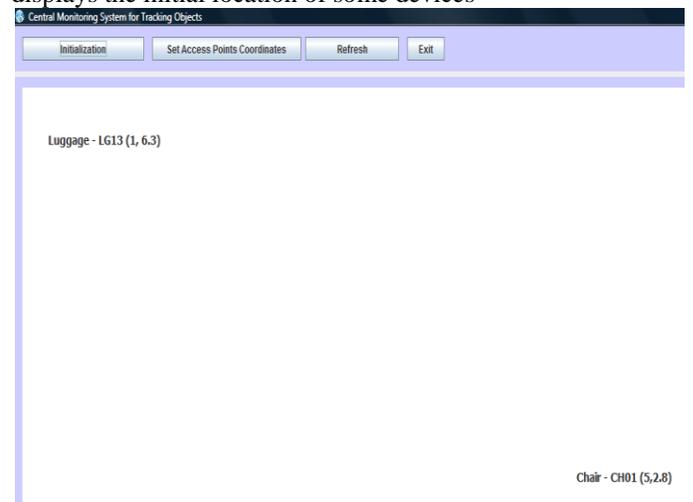

Fig. 11. Initial Location of Some Devices

When a device is moved, the Central Monitoring System sounds alarm and the current location is displayed. The black colour of a moving object being tracked is changed to red. The figure below illustrates the movement of a device being tracked.

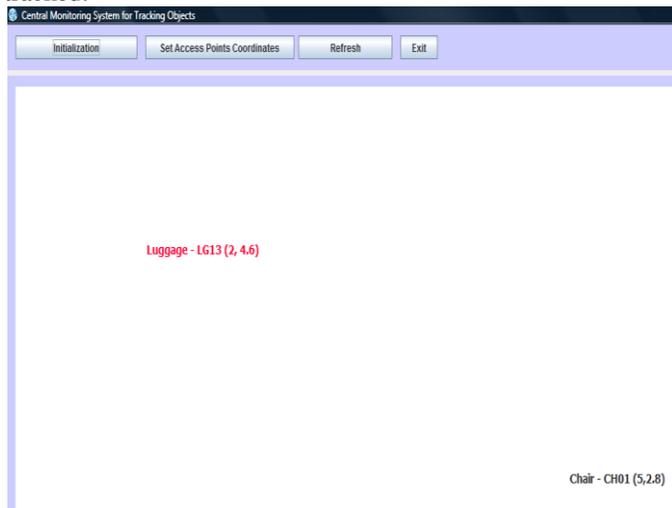

Fig. 12. Illustration of a Moved Object Being Tracked

When the Refresh button is clicked, the alarm stops sounding and the red coloured object is changed to the default colour (black). The Central Monitoring System can only be closed, by clicking on the Exit button or the closed button, when all the connecting devices have been disconnected from it.

## V. SYSTEM ANALYSIS

The system implementation was successful. However, some activities that usually affected system's performance are discussed in this section

### A. When an Access Point is Disconnected

The performance of the system is affected adversely when an Access Point is disconnected from the device being tracked. Wrong coordinates are usually estimated resulting in false alarm generation. To prevent such false alarm generation, an error message is displayed on the screen of the Central Monitoring System immediately an Access Point is disconnected.

### B. Effect of Power Consumption

There is an excessive power consumption associated with Bluetooth technology which usually affects system's performance. It is therefore suitable to implement the Access Points as J2SE systems on PCs instead of J2ME systems. The excessive power consumption quickly runs batteries of J2ME systems down resulting in frequent system initializations

### C. Intervals between Floor Measurement or XY Points

The smaller the interval of measurement or points in the XY plane, the greater the margin of errors and hence the numerous generation of false alarm. It was deduced that good system performance was achieved when the intervals between points in the XY plane were 3 meters to 5 meters

## VI. CONCLUSION

A new concept for tracking indoor objects has been proposed. The object being tracked is initialized using least square statistical method whereas principles of motion supported by Bluetooth Communication technology are used to track the object. Although the system is computationally intensive for the Central Monitoring System yet the Central Monitoring System being a powerful system handles computations effectively and efficiently. The system serves as a basis for implementing tracking system that has the following characteristics to enhance its performance and functionalities:

- It does not require line-of-sight operation.
- The tracker system and the Access Point system implementations require low memory and computational overload.

REFERENCES

[1] B. Antic, J. O. N. Castaneda, D. Culibrk, A. Pizurica, V. Crnojevic and W. Philips, "*Robust Detection and Tracking of Moving Objects in Traffic Video Surveillance*", Advanced Concepts for Intelligent Vision Systems, pages 494-505, 2009
[2] R. Bajaj, S.L. Ranaweera and D.P. Agrawal, "*GPS: Location-Tracking Technology*", IEEE Vol.35, No.4, p. 92–94, 2002
[3] R. Want, "*An Introduction to RFID Technology*", IEEE Pervasive Computing, p. 25-33, 2006
[4] C. Dawson, "*Device Tracking on a Scattered Bluetooth-Enabled Network*", Bsc Dissertation, Faculty of Engineering, University of Bristol, May 2005
[5] A. Kivimaki, V. Fomin, "*What Makes a Killer Application for the Cellular Telephony Services?*", 2nd IEEE Conference on Standardization and Innovation in Information Technology, p. 25-37, 2001
[6] B. Ghribi, L. Logrippo, "*Understanding GPRS: The GSM Packet Radio Service*", School of Information Technology and Engineering, University of Ottawa
[7] Z. Sheng and J.K. Pollard, "*Position Measurement using Bluetooth*" IEEE Transactions on Consumer Electronics, Vol 52, Issue 2, p. 555 – 558, 2006.
[8] Bluetooth Special Interest Group Specification, "*Specification of the Bluetooth System Core*" Vol 1 and Vol 2, Versions 1.1, February 22, 2001
[9] P. Bahl and V. Padmanabhan, "*Radar: An in-building RF_based* User Location *and Tracking System*", Proceedings of the IEEE Infocom 2000, Tel-Aviv, Israel, vol. 2, p. 775-784, Mar. 2000.
[10] S. Rathi, "*Infrastructure: Bluetooth Protocol Architecture*", Microware Architect, Microware System Corporation, p. 1 – 6, 2000
[11] B. Hopkins and R. Antony, "*Bluetooth for Java*", ISBN 1-59059-78-3, pp 33-35, 2003.
[12] A. Zapater, K. Kyamakya, S. Feldmann, M. Kruger, I. Adusei, "*Development and Implementation of a Bluetooth Networking Infrastructure for a Notebook*", University Scenario, International Conference on Wireless Networks, 2003.
[13] S. K. Opoku, "*Performance Enhancement of Large-Size NFC Multi-Touch System*", Cyber Journals: Multidisciplinary Journals in Science and Technology, Journal of Selected Areas in Telecommunications (JSAT), Vol. 2, No. 10 October Edition, 2011
[14] S. Beji, E. N. Kadhi, "*An Overview of Mobile Applications Architecture and the Associated Technologies*", Fourth International Conference on Wireless and Mobile Communications, ICWMC '08, IEEE, p. 77 – 83, 2008

[15] M.J. Yuan, "*Entreprise J2ME, Developing Mobile Java Applications*", Upper Saddle River: Prentice Hall PTR, 2006, pp. 20-25
[16] Foundation Profile Overview
http://java.sun.com/products/foundation/overview.html
[17] JSR 219, Foundation Profile Specification
http://www.jcp.org/en/jsr/detail?id=219
[18] Personal Basis Profile Overview
http://java.sun.com/products/personalbasis/overview.html
[19] JSR 217, Personal Basis Profile Specification
http://www.jcp.org/en/jsr/detail?id=217
[20] Personal Profile Overview
http://java.sun.com/products/personalprofile/overview.html
[21] JSR 62, Personal Profile Specification
http://www.jcp.org/en/jsr/detail?id=62
[22] L. Doug, "Java *All-in-One Desk Reference for Dummies*", Wiley Publishing Inc, ISBN-13: 978-0-7645-8961-4, pg 65-70, 2005
[23] M. M. Organero, S. K. Opoku, "*Using NFC Technology for Fast-Tracking Large-Size Multi-Touch Screens*", Cyber Journals: Multidisciplinary Journals in Science and Technology, Journal of Selected Areas in Telecommunications (JSAT), Vol. 2, No. 4 April Edition, 2011
[24] P. Dawkins, "*Linear Algebra, Vector Spaces of College Algebra*", p.106-113, 2007 Available: http://tutorial.math.lamar.ed/terms.aspx